\documentclass[a4paper,fleqn,usenatbib]{mnras}
\usepackage[T1]{fontenc}
\usepackage{ae,aecompl}
\usepackage{graphicx}
\usepackage{multirow}
\usepackage{caption,hyperref}

\def\jnl@style{\it}
\def\aaref@jnl#1{{\jnl@style#1}}

\def\aaref@jnl#1{{\jnl@style#1}}

\def\aj{\aaref@jnl{AJ }}                   
\def\araa{\aaref@jnl{ARA\&A }}             
\def\apj{\aaref@jnl{ApJ }}                 
\def\apjl{\aaref@jnl{ApJ }}                
\def\apjs{\aaref@jnl{ApJS }}               
\def\ao{\aaref@jnl{Appl.~Opt. }}           
\def\apss{\aaref@jnl{Ap\&SS }}             
\def\aap{\aaref@jnl{A\&A }}                
\def\aapr{\aaref@jnl{A\&A~Rev. }}          
\def\aaps{\aaref@jnl{A\&AS }}              
\def\azh{\aaref@jnl{AZh }}                 
\def\baas{\aaref@jnl{BAAS }}               
\def\jrasc{\aaref@jnl{JRASC }}             
\def\memras{\aaref@jnl{MmRAS }}            
\def\mnras{\aaref@jnl{MNRAS }}             
\def\pra{\aaref@jnl{Phys.~Rev.~A }}        
\def\prb{\aaref@jnl{Phys.~Rev.~B }}        
\def\prc{\aaref@jnl{Phys.~Rev.~C }}        
\def\prd{\aaref@jnl{Phys.~Rev.~D }}        
\def\pre{\aaref@jnl{Phys.~Rev.~E }}        
\def\prl{\aaref@jnl{Phys.~Rev.~Lett. }}    
\def\pasp{\aaref@jnl{PASP }}               
\def\pasj{\aaref@jnl{PASJ }}               
\def\qjras{\aaref@jnl{QJRAS }}             
\def\skytel{\aaref@jnl{S\&T }}             
\def\solphys{\aaref@jnl{Sol.~Phys. }}      
\def\sovast{\aaref@jnl{Soviet~Ast. }}      
\def\ssr{\aaref@jnl{Space~Sci.~Rev. }}     
\def\zap{\aaref@jnl{ZAp }}                 
\def\nat{\aaref@jnl{Nature }}              
\def\iaucirc{\aaref@jnl{IAU~Circ. }}       
\def\aplett{\aaref@jnl{Astrophys.~Lett. }} 
\def\apspr{\aaref@jnl{Astrophys.~Space~Phys.~Res. }}
\def\bain{\aaref@jnl{Bull.~Astron.~Inst.~Netherlands }} 
\def\fcp{\aaref@jnl{Fund.~Cosmic~Phys. }}  
\def\gca{\aaref@jnl{Geochim.~Cosmochim.~Acta }}   
\def\grl{\aaref@jnl{Geophys.~Res.~Lett. }} 
\def\jcp{\aaref@jnl{J.~Chem.~Phys. }}      
\def\jgr{\aaref@jnl{J.~Geophys.~Res. }}    
\def\jqsrt{\aaref@jnl{J.~Quant.~Spec.~Radiat.~Transf. }}
\def\memsai{\aaref@jnl{Mem.~Soc.~Astron.~Italiana }}
\def\nphysa{\aaref@jnl{Nucl.~Phys.~A }}   
\def\physrep{\aaref@jnl{Phys.~Rep. }}   
\def\physscr{\aaref@jnl{Phys.~Scr }}   
\def\planss{\aaref@jnl{Planet.~Space~Sci. }}   
\def\procspie{\aaref@jnl{Proc.~SPIE }}   

\newcommand{\vpfit}{\texttt{VPFIT}}
\newcommand{\da}{$\Delta\alpha / \alpha$} 
\defcitealias{whitmore:2014}{WM}
\title[Distortion modelling]{Modelling long-range wavelength distortions
in quasar absorption echelle spectra}
      
\author[V. Dumont, J. K. Webb]
       {V. Dumont$^{1,2}$\thanks{E-mail:vincentdumont11@gmail.com}, J. K. Webb$^{1}$\\
         $^{1}$School of Physics, University of New South Wales, Sydney NSW 2052, Australia\\
         $^{2}$Department of Physics, University of California, Berkeley, California 94720-7300, USA}

\begin{document}

\date{Accepted 2017 February 9. Received 2017 February 7; in original form 2016 December 20}

\pagerange{\pageref{firstpage}--\pageref{lastpage}} \pubyear{2016}

\maketitle

\label{firstpage}

\begin{abstract}
Spectra observed with the Ultraviolet and Visual Echelle Spectrograph
(UVES) on the European Southern Observatory's VLT exhibit long-range
wavelength distortions. These distortions impose a systematic error on
high-precision measurements of the fine-structure constant, $\alpha$,
derived from intervening quasar absorption systems. If the distortion
is modelled using a model that is too simplistic, the resulting bias
in \da\ away from the true value can be larger than the statistical
uncertainty on the $\alpha$ measurement. If the effect is ignored altogether,
the same is true. If the effect is modelled properly, accounting for the
way in which final spectra are generally formed from the co-addition of
exposures made at several different instrumental settings, the effect
can be accurately removed and the correct \da\ recovered.
\end{abstract}

\begin{keywords}
cosmological parameters -- techniques: spectroscopic --
methods: data analysis -- quasars: absorption lines
\end{keywords}

\section{Introduction}

\subsection{Measuring the fine-structure constant}

The Many Multiplet method permits precise measurements of the fine
structure constant $\alpha$ using absorption systems in high-resolution
quasar spectra \citep{webb:1999}. The method has been used extensively
to study possible space-time variation of alpha in the Universe. The
largest sample to date \citep{king:2012} comprises 154 measurements
obtained from spectra taken with the UVES spectrograph on the VLT
telescope in Chile, combined with 143 earlier measurements made with the
HIRES spectrograph mounted on the Keck telescope in Hawaii
\citep{murphy:2003,murphy:2004}. The extensive sky coverage of that
large sample permitted the first accurate constraints on any possible
spatial variation of $\alpha$ over cosmological scales.  A tentative
detection of spatial variation was reported in \citet{webb:2011} and \citet{king:2012} with a
statistical significance of 4.1$\sigma$, allowing for both statistical
and systematic uncertainties. The systematic uncertainties in that
analysis were estimated as free parameters so did not rely on
identifying and quantifying specific systematics. Long-range wavelength
distortions in echelle spectrographs had not been measured at that time
so were not taken into account explicitly in \citet{webb:2011} and \citet{king:2012}.

\subsection{Searching for long-range wavelength distortions}

\citet{2008A&A...481..559M} first searched for possible wavelength
distortions in high-resolution quasar spectra by correlating the
reflected solar spectrum from asteroid observations observed using UVES
with absolute solar calibrations. That study found no evidence for
long-range wavelength distortion for VLT/UVES spectra.\\

Subsequently, \citet{rahmani:2013} used the same method but
obtained a grater precision and showed
that in fact long-range wavelength distortions do occur in UVES spectra
and that, for a single exposure, the form of the distortion appears to
be reasonably well approximated by a simple linear function of velocity
shift versus observed wavelength.\\

More recently, \cite{whitmore:2014}, hereafter WM, made further measurements
of the long-range wavelength distortion effect in UVES spectra, confirming
the linear trends reported in \citet{rahmani:2013}. \citetalias{whitmore:2014}
then attempted to estimate the impact of this effect on the analysis of
\citet{king:2012} by applying the simple linear long-range distortion
function seen in a single asteroid or solar twin exposure.

\subsection{The danger of mis-modelling}

Since quasars are generally rather faint objects, multiple exposures are
typically made in order to obtain a sufficiently good signal to noise
ratio. Different central wavelength settings are generally used in order
to end up with a final spectrum spanning the visual wavelength range.
The vast majority of UVES archival quasar spectra have been observed this
way, that is, a final co-added quasar spectrum is formed from exposures
taken at many different wavelength settings. For example, of the 154
\da\ measurements reported in \citet{king:2012}, only 12 (or 8\%) 
were observed at a single wavelength setting.\\

An interesting and systematic characteristic of the long-range
distortions seen in the asteroid or solar-twin measurements (which are
single exposures) is that the zero-point (i.e. the wavelength at which
there is zero distortion) appears to coincide with the central
wavelength of the exposure (see e.g. Figure 7 in \citet{rahmani:2013}
and Figure 4 in \citetalias{whitmore:2014}). In the simulations
described in the present paper, we adopt this same characteristic, but
note that the model does not permit a constant velocity offset between
different exposures that contribute to a co-added quasar spectrum.  We
shall address this point explicitly in a separate paper.\\

Clearly this means that the resulting long-range distortion function for
any given quasar spectrum should be described by an appropriate
co-addition of the distortion functions corresponding to each individual
quasar exposure. Nevertheless, \citetalias{whitmore:2014} applied
corrections to the co-added quasar sample reported in
\citet{webb:1999} using a distortion model derived from a
single exposure of an asteroid measurement and used the results to
concluded that such a distortion was able to explain the $\alpha$ dipole
signal previously reported.\\

We shall explicitly address the specific impact of long-range wavelength
distortions on the \citet{king:2012} sample in a separate paper. The aim
of the present paper is to illustrate, using a case study, the
importance of deriving an appropriate correction function.  We show that
applying a simplistic model to a quasar spectrum does not have the
effect of ``correcting'' any actual distortion, but instead has the
effect of {\it introducing} a spurious distortion and hence biasing any
estimate of $\alpha$.\\

The remainder of this paper is structured as follows: In Section
\ref{sec:wavecorr}, we will show how to determine the distortion
function for each co-added spectrum used for \da\ measurements. Then,
using simulations, in Section \ref{simulation} we show how to use the
quasar spectrum itself to solve for the distortion function and hence
derive a correction to \da, given some simplifying assumptions.
In Section \ref{recovery}, we compare our distortion modelling with the
simplistic model used in \citetalias{whitmore:2014} and show that using
the wrong model produces
the wrong answer, that is, one can end up with a spurious estimate of
both the distortion and the estimated value of \da, quantifying the
impact using numerical simulations of one particular quasar spectrum.

\section{Modelling long-range distortion} \label{sec:wavecorr}

\subsection{Assumptions made}

In this paper, we make the following assumptions concerning the
distortion pattern: (1) the distortion is linear in observed wavelength,
with a zero point at the central wavelength of each exposure, and (2) we
adopt a constant slope for the linear distortion pattern for all
exposures on the same quasar spectrum.\\

The first of these assumptions follows from inspection of the asteroid
and solar twin measurements from \citetalias{whitmore:2014} and others.
Although the asteroid and solar twin measurements are seen to vary from
observation to observation, our second assumption infers we are taking a
mean value of the slope over all observations contributing to the final
co-added quasar spectrum. We will quantify the consequences of the
second assumption in a separate paper.

\begin{figure}
  \centering
  \includegraphics[width=\columnwidth]{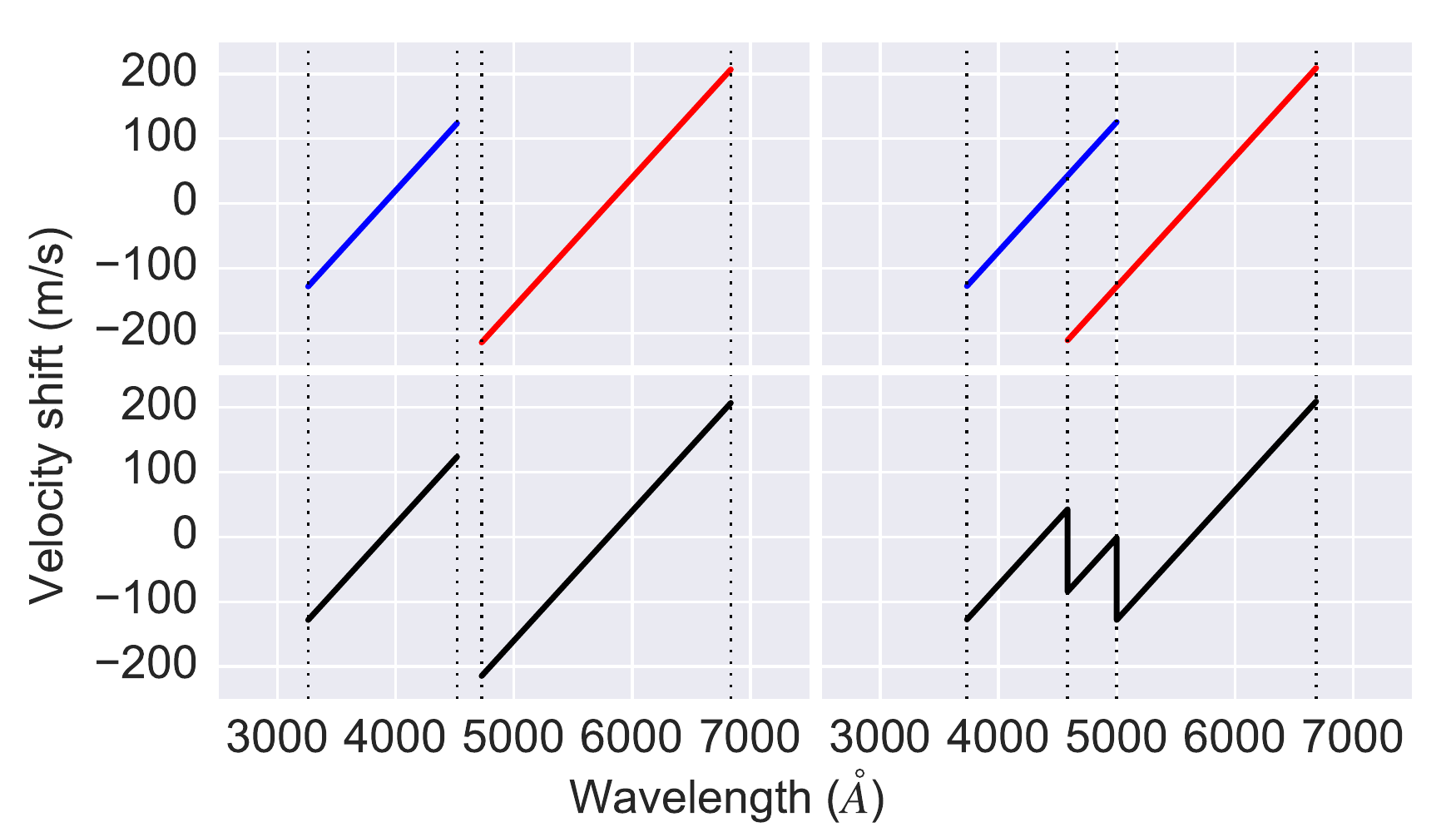}
  \caption{Illustration of the distortion patterns corresponding to an
  example set of quasar exposures. The top panels show the individual
  distortion models for two science exposures with different wavelength
  settings for the UVES spectrograph on the VLT. The top left panel
  corresponds to the 390/580 nm wavelength settings and the top right
  panel correspond to the 437/564 nm settings. The bottom panels show
  the resulting distortion function. When no overlaps occur between
  exposures, the final distortion function results in a discontinuous
  model where regions outside the wavelength range covered by the
  exposures are simply ignored. On the other hand, if there is overlap
  between exposures, this will produce the saw-tooth model seen in the
  bottom right-hand panel.}
  \label{fig:curves}
\end{figure}

\subsection{Calculating the distortion function}

For a single science exposure, the velocity distortion function for
that exposure (the $i^{th}$ exposure), is:

\begin{equation}
\label{distsingle}
v_{\mathrm{dist,i}}(\lambda)=\gamma\left(\lambda-\lambda_{\mathrm{cent,i
}}\right)
\end{equation}

where $\gamma$ is the slope of the linear distortion model,
and $\lambda_{\mathrm{cent,i}}$ is the central wavelength of the science
exposure.\\

In general, each spectrum is formed by combining multiple science
exposures taken at different wavelength settings.  In this case the
composite distortion pattern therefore depends on the central wavelength
and the wavelength edges of each exposure (Fig. \ref{fig:curves}). In
order to correctly estimate the actual velocity distortion at a given
wavelength, one needs to take into account the details of all exposures
contributing to that wavelength.\\

We denote $\lambda_{\mathrm{start,i}}$, $\lambda_{\mathrm{end,i}}$ be the
start and end wavelength of a given science exposure of index $i$. The
net velocity shift $v_{\mathrm{net}}(\lambda)$ at a given wavelength,
$\lambda$, has contributions from exposures satisfying:

\begin{equation}
\label{expconsider}
\lambda_{\mathrm{start,i}} < \lambda < \lambda_{\mathrm{end,i}}
\end{equation}

The net distortion shift in the final co-added spectrum depends on the
signal to noise ratio of each contributing science exposure. We
therefore form the weighted net distortion function using weighting
factors proportional to the square root of the exposure time for the
$i^{th}$ exposure,

\begin{equation}
\label{distfinal}
v_{\mathrm{net}}(\lambda)=\frac{\sum\limits_{i}\left(\sqrt{T_i}
v_{\mathrm{dist,i}}(\lambda)\right)}{\sum\limits_{i}\sqrt{T_i}}
\quad\textrm{where}\quad \lambda_{\mathrm{start,i}} < \lambda <
\lambda_{\mathrm{end,i}}
\end{equation}

Assuming the slope of the distortion function to be the 
same for every science exposures (we shall address this approximation
in a separate paper), the net velocity distortion function is:

\begin{equation}
\label{eq:vfinal}
v_{\mathrm{net}}(\lambda)=\frac{\gamma\sum\limits_{i}\left(\sqrt{T_i}
\left(\lambda-\lambda_{\mathrm{cent,i}}\right)\right)}{\sum\limits_
{i}\sqrt{T_i}}
\end{equation}

\section{Synthetic spectra}
\label{simulation}

\begin{figure}
\centering
\includegraphics[width=\columnwidth]{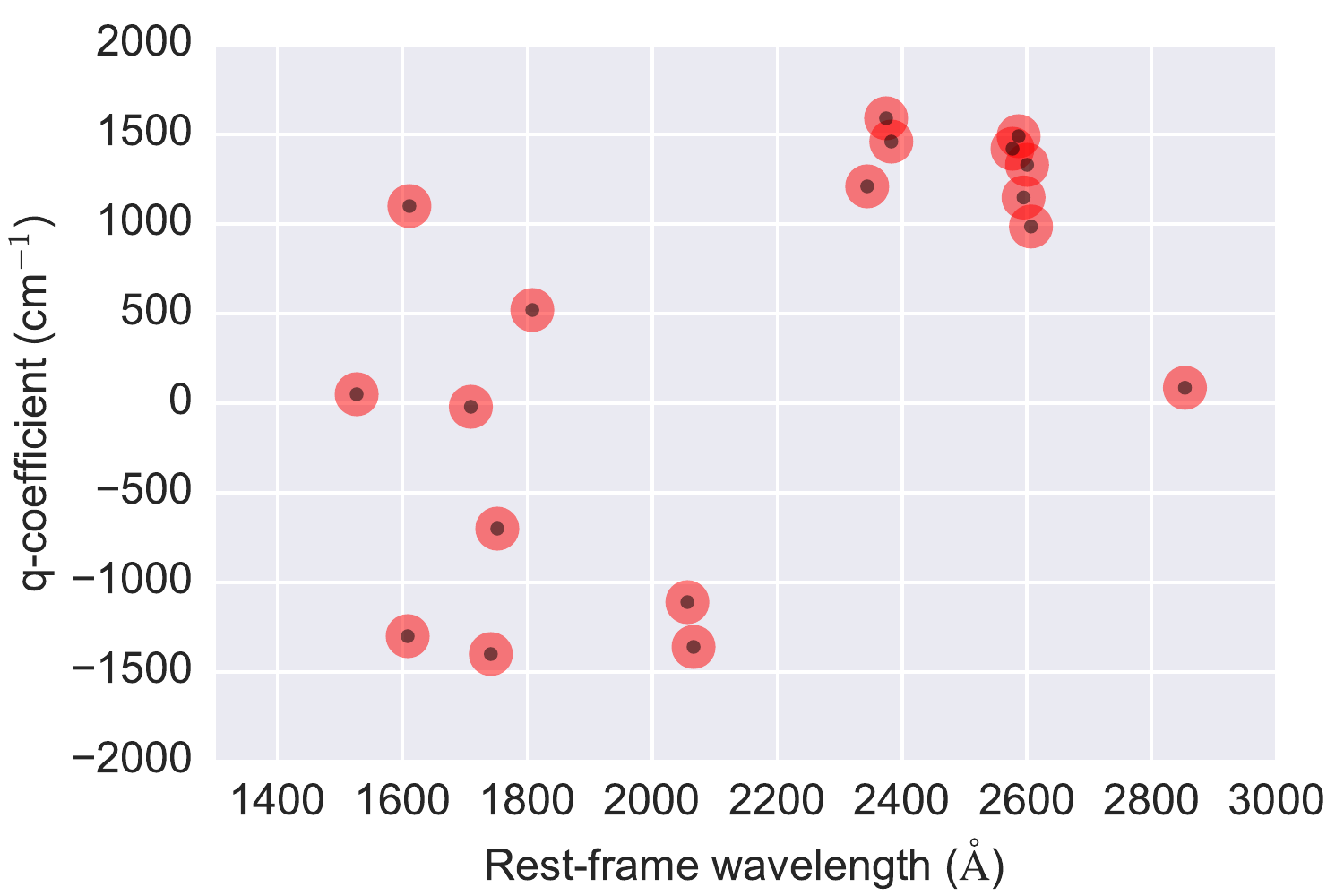}
\caption{Sensitivity to alpha against rest-wavelength for each transition
used in the simulated spectrum. The q-coefficients and rest-wavelengths
can be found in Table \ref{table:reglist}. }
\label{qscatter}
\end{figure}

Figure \ref{qscatter} illustrates the $\alpha$ sensitivity coefficients,
$q$, for the transitions detected in the $z_\mathrm{abs}=1.3554$ absorption
system towards the $z_\mathrm{em}=1.94$ quasar J043037-485523.  The overall
range in $q$ is $\sim 3000$. The points are widely scattered and do not
correlate tightly with rest-frame wavelength.  It is this property that
breaks degeneracy between the parameters \da\ and $v_\mathrm{net}(\lambda)$ and
allows us to solve explicitly for both parameters simultaneously.\\

We have generated a simulated spectrum of the $z_\mathrm{abs}=1.3554$
absorption system towards J043037-485523, using a signal to noise per
pixel of 1000, a pixel size of 2.5 km/s, and a Gaussian instrumental
resolution of $\sigma = 2.55$ km/s. The latter two parameters match
those of actual data used previously for a measurement of \da,
\citep{king:2012}.\\

We choose a real absorption system for this simulation to emulate
reality as far as possible, albeit at very high signal to noise ratio.
The useful characteristics of this system are:
(1) it has a complex velocity structure, like most
systems used to derive stringent constraints on \da,
(2) it exhibits a large number of transitions (20 in total), 
including transitions with high sensitivity to \da, and 
(3) the velocity structure and other line parameters for this 
system are such that it yields a stringent constraint on the
measurement of \da, ($-4.05\pm2.32$ ppm, \citet{king:2012}).

\subsection{Methodology}

Our aim is to explore the impact on estimating \da\ using two
types of distortion models to correct the spectrum. To do this we generate a
simulated spectrum with known \da\ (=0) and upon which has been imposed a
sawtooth distortion model of the type illustrated in the middle
hand panel of Figure \ref{fig:distmodel}. That spectrum is then modelled in
two ways: using the \citetalias{whitmore:2014} model comprising one single
linear distortion function for each arm of the UVES spectrograph
(illustrated in the bottom panel of Figure \ref{fig:distmodel}) and using the 
input (and more correct) sawtooth form.\\

The synthetic spectrum is created with random noise added, using
signal to noise of 1000.  We used a high S/N for these simulations
to provide the accurate measurements needed to properly illustrate 
the importance of using a more realistic distortion model.
Although S/N=1000 corresponds to an unrealistically
high signal-to-noise ratio for a real quasar echelle spectrum at
the present time, it may be achievable using future generations 
of large telescopes.\\

Throughout the fitting procedure, the slope of the distortion function 
is treated as a free parameter which we solve for simultaneously with 
the other ``interesting'' parameter, \da, using \vpfit\footnote{Version 10,
Carswell, R.F. and Webb, J.K., \vpfit{} - Voigt profile fitting program
\url{http://www.ast.cam.ac.uk/~rfc/vpfit.html}}.  In practice, 
the calculations were carried out in the following sequence: 
$\gamma = 0.2 \rightarrow 0.25$, then $\gamma = 0.2 \rightarrow 0$, 
in steps of $\delta\gamma=0.005$.  At each successive step after the
first, the starting parameter guesses supplied to \vpfit{} are the 
best-fit results from the previous step.  At the two starting points,
the correct (known) parameters values are used as starting guesses.  
The sequence and method above should make no difference to the final 
results (if it were done in a different sequence, we would get the same
answer) but we provide the details here for completeness.

\subsection{Absorption line parameters for the synthetic spectrum}

The absorption line parameters used in generating the synthetic spectrum
come from \citet{king:2012}. However, we removed the AlII 1670 and FeII
2260 transitions from the original model for the following reasons: (1)
AlII 1670 was excluded because it is saturated and hence adds little in
terms of sensitivity to \da; (2) FeII 2260 was excluded because the
isotopic structure is unknown (this is not the case for the other FeII
transitions used in our model).  Table \ref{table:complist} gives the
line parameters used to create the synthetic spectrum. The b-parameters
are taken as turbulent rather than thermal.  We impose on the synthetic
spectrum a sawtooth distortion model with a slope of +0.2 m/s/\AA. This
value is within the range of typical values measured in asteroid or
solar-twin spectra.  The aim is to see how well we can recover the input
slope value.

\begin{table*}
  \caption{Parameter first guesses for each component fitted in the
  system. The first 2 columns show for each component the redshift $z$,
  Doppler parameter $b$ while the remaining 6 columns show the column
  density $N$ of each ion at each redshift location.}
  \begin{tabular}{ccccccccc}
    \hline\hline \\ [-2.0ex]
    \multirow{2}{*}{$z$} & $b$ & $\log(N_\mathrm{MgI})$ & $\log(N_\mathrm{SiII})$ &
    $\log(N_\mathrm{CrII})$ & $\log(N_\mathrm{MnII})$ &
    $\log(N_\mathrm{NiII})$ & $\log(N_\mathrm{FeII})$ \\ [0.8ex]
    &[km/s]&[cm$^{-2}$]&[cm$^{-2}$]&[cm$^{-2}$]&[cm$^{-2}$]&[cm$^{-2}$]&[cm$^{-2}$]\\
    \hline \\ [-2.0ex]
    1.3550877 & 4.6450 & 11.04882 & 13.24512 &        - &        - & 12.08916 & 12.86596\\
    1.3551633 & 4.2369 & 11.01670 & 13.38996 &        - & 11.17501 &        - & 13.15444\\
    1.3552067 & 4.1203 & 11.43565 & 13.52086 & 10.82803 & 11.12017 & 12.51781 & 13.13221\\
    1.3552588 & 3.1190 & 11.16424 & 13.99763 & 12.18494 & 11.54568 & 12.52872 & 13.96588\\
    1.3553108 & 2.0405 & 10.71039 & 13.82403 & 11.91851 & 11.43706 & 12.28327 & 13.59766\\
    1.3553808 & 5.5259 & 11.37225 & 14.33999 & 12.39546 & 11.80985 & 12.86060 & 14.17853\\
    1.3554579 & 1.9351 &        - & 13.20139 &        - & 11.04225 &        - & 12.99883\\
    1.3555581 & 7.0834 & 10.88266 & 14.28980 & 12.39132 & 11.61898 & 12.78850 & 14.01932\\
    1.3555835 & 1.9147 & 10.41418 & 13.75303 & 11.92118 & 11.45834 & 12.03819 & 13.72489\\
    \hline \\ [-2.0ex]
  \end{tabular}
  \label{table:complist}
\end{table*}

\subsection{Distortion model}

\begin{figure}
\centering
\includegraphics[width=\columnwidth]{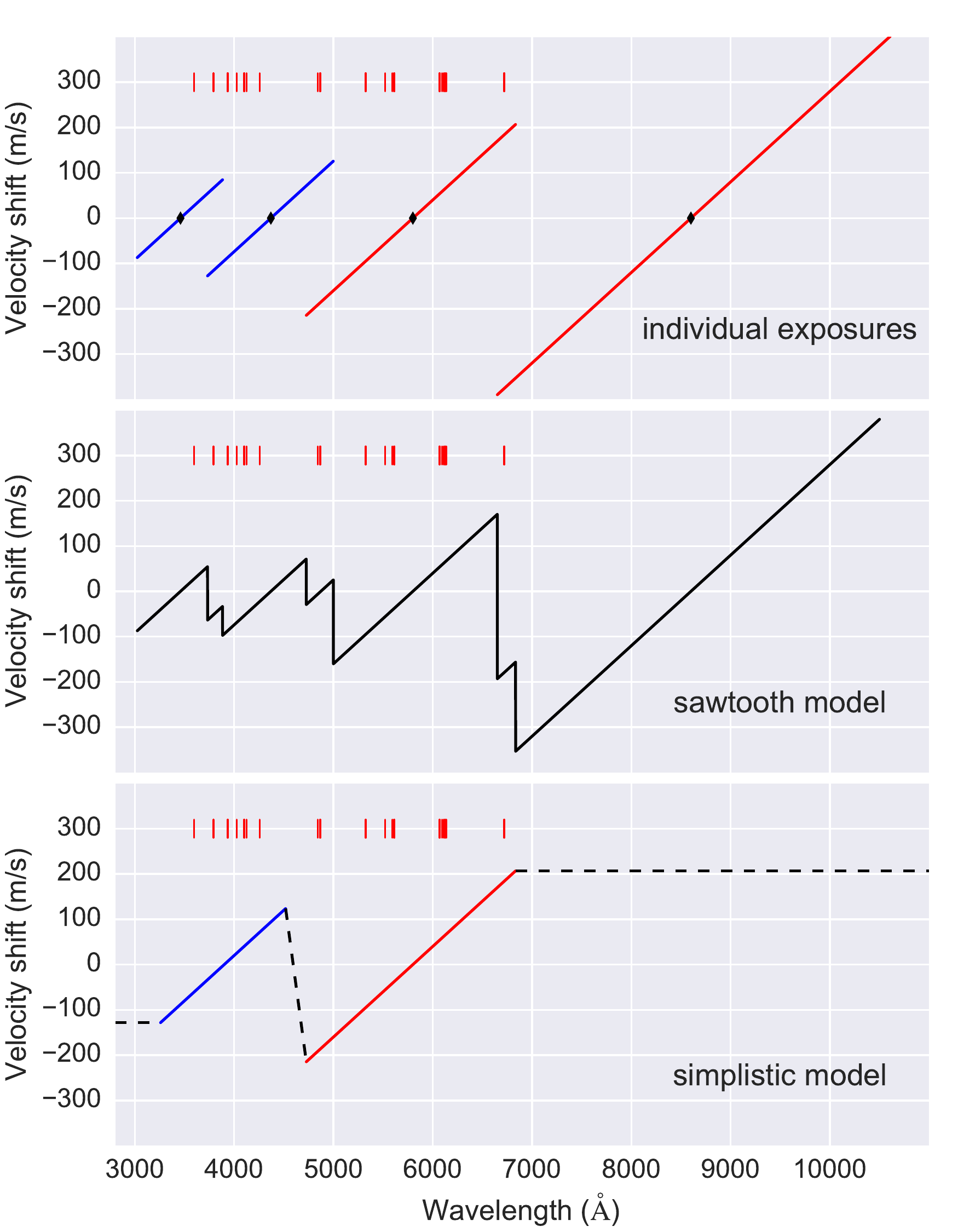}
\caption{Long-range distortion model of J043037-485523 for a distortion
  slope of +0.2 m/s/\AA. The top panel shows the distortion pattern for 
  each individual science exposure present in our sample. The middle
  panel shows the combined distortion model. The simplistic distortion
  model as used by \citetalias{whitmore:2014} is presented in the bottom
  panel, where the blue (solid line on the left) and red (solid line on
  the right) lines represent the long-range distortion pattern for the blue
  and red arms of the UVES spectrograph. The extend of the solid lines are
  from \citetalias{whitmore:2014}. The distortion outside the range indicated
  by the solid lines is assumed by \citetalias{whitmore:2014} to be constant and
  is represented by horizontal dashed lines. The region between both arm models 
  are simply connected with a straight line although there seems to be no
  empirical justification for this. The red tick-marks on every panel
  illustrate the wavelength of each transition fitted in the system. }
  \label{fig:distmodel}
\end{figure}

\begin{figure*}
\centering
\includegraphics[width=\textwidth]{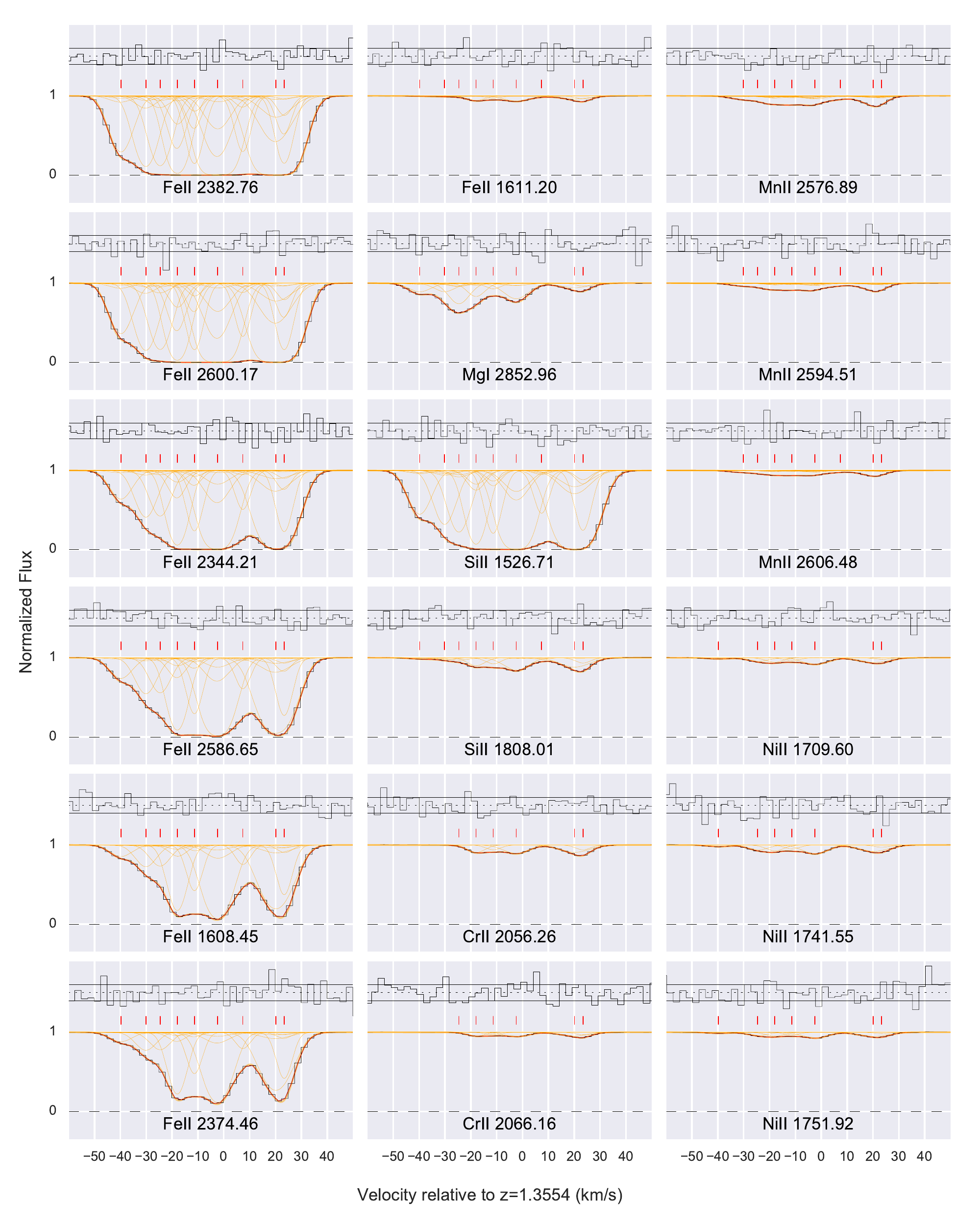}
\caption{Velocity plot of each simulated fitting region. The \vpfit{}
model is illustrated in red while each individual Voigt profile from
each component and transition is in orange. The normalised residuals are
plotted at the top of each panel.}
\label{voigt_profiles}
\end{figure*}

We tabulate in Table \ref{table:explist} all the science exposures used
to produce the combined spectrum of J043037-485523. The wavelength edges
of each exposure were recovered using the UVES Exposure Time Calculator
\footnote{\url{http://www.eso.org/observing/etc/bin/gen/form?INS.NAME=UVES+INS.MODE=spectro}}.
The distortion model J043037-485523 can then be built by applying
Eq. \ref{eq:vfinal} on the ensemble of exposures (Fig. \ref{fig:distmodel}).
One can therefore estimate, for each transition, the value of the shift
due to long-range distortion effect in the spectrum and which can be
applied to the simulated spectrum to account for such distortion. 
Table \ref{table:reglist} tabulates the transitions fitted,
the exposures covering each region, and the
velocity shift corresponding to a distortion slope of +0.2 m/s/\AA.
Figure \ref{voigt_profiles} shows the final simulated and distorted
absorption system.

\begin{table*}
  \caption{List of 18 individual science exposures for J043037-485523
  in order of central wavelength combined to form the final co-added
  spectrum. A total of 4 different wavelength settings have been used,
  centered at 346, 437, 580, and 860 nm.}
  \begin{tabular}{rccllccccr}
    \hline\hline \\ [-2.0ex]
    \multirow{2}{*}{\#} &
    \multirow{2}{*}{Dataset ID} &
    $T_{\mathrm{exp}}$ &
    \multirow{2}{*}{arm} &
    \multirow{2}{*}{mode} &
    \multirow{2}{*}{grating} &
    $\lambda_{\mathrm{start}}$ &
    $\lambda_{\mathrm{cent}}$ &
    $\lambda_{\mathrm{end}}$ \\
    & & [second] & & & & [nm] & [nm] & [nm]\\
    \hline \\ [-2.0ex]
    1  & UVES.2001-02-01T02:05:23.054 & 3600 & BLUE & DICHR\#1 & CD\#1 & 302.45 & 346 & 388.40 \\
    2  & UVES.2001-03-18T00:03:50.906 & 3600 & BLUE & DICHR\#1 & CD\#1 & 302.45 & 346 & 388.40 \\
    3  & UVES.2001-03-19T00:15:14.848 & 3600 & BLUE & DICHR\#1 & CD\#1 & 302.45 & 346 & 388.40 \\
    4  & UVES.2001-01-13T01:28:38.684 & 1434 & BLUE & DICHR\#2 & CD\#2 & 373.24 & 437 & 499.94 \\
    5  & UVES.2001-01-16T02:59:19.308 & 2922 & BLUE & DICHR\#2 & CD\#2 & 373.24 & 437 & 499.94 \\
    6  & UVES.2001-02-01T03:11:26.409 & 3600 & BLUE & DICHR\#2 & CD\#2 & 373.24 & 437 & 499.94 \\
    7  & UVES.2001-02-14T02:48:10.272 & 3600 & BLUE & DICHR\#2 & CD\#2 & 373.24 & 437 & 499.94 \\
    8  & UVES.2001-03-05T00:19:43.715 & 3600 & BLUE & DICHR\#2 & CD\#2 & 373.24 & 437 & 499.94 \\
    9  & UVES.2001-03-06T00:14:46.378 & 3600 & BLUE & DICHR\#2 & CD\#2 & 373.24 & 437 & 499.94 \\
    10 & UVES.2001-02-01T02:05:21.080 & 3600 & RED  & DICHR\#1 & CD\#3 & 472.69 & 580 & 683.49 \\
    11 & UVES.2001-03-18T00:03:49.697 & 3600 & RED  & DICHR\#1 & CD\#3 & 472.69 & 580 & 683.49 \\
    12 & UVES.2001-03-19T00:15:14.827 & 3600 & RED  & DICHR\#1 & CD\#3 & 472.69 & 580 & 683.49 \\
    13 & UVES.2001-01-13T01:28:37.206 & 1437 & RED  & DICHR\#2 & CD\#4 & 665.06 & 860 & 060.57 \\
    14 & UVES.2001-01-16T02:59:21.364 & 2921 & RED  & DICHR\#2 & CD\#4 & 665.06 & 860 & 060.57 \\
    15 & UVES.2001-02-01T03:11:21.253 & 3600 & RED  & DICHR\#2 & CD\#4 & 665.06 & 860 & 060.57 \\
    16 & UVES.2001-02-14T02:48:08.997 & 3600 & RED  & DICHR\#2 & CD\#4 & 665.06 & 860 & 060.57 \\
    17 & UVES.2001-03-05T00:19:40.291 & 3600 & RED  & DICHR\#2 & CD\#4 & 665.06 & 860 & 060.57 \\
    18 & UVES.2001-03-06T00:14:41.535 & 3600 & RED  & DICHR\#2 & CD\#4 & 665.06 & 860 & 060.57 \\
    \hline \\ [-2.0ex]
  \end{tabular}
  \label{table:explist}
\end{table*}

\begin{table*}
  \centering
  \caption{Parameters for each fitting region in the synthetic
  spectrum. The first three columns specify the ion, laboratory
  rest-wavelength and q-coefficient for each transition. The next
  three columns correspond to the first, middle and last observed
  wavelengths for each region. The second-last column shows the
  range of exposures tabulated in Table \ref{table:explist} that
  covers each region of interest. Finally, the rightmost column
  gives the expected velocity shift at each observed central
  wavelength due to distortion effect when a slope of 0.2 m/s is
  applied to the distortion model.}
  \begin{tabular}{lcrccccr}
    \hline\hline \\ [-2.0ex]
    \multirow{2}{*}{Ion} & $\lambda_\mathrm{lab}$ & q &
    $\lambda_\mathrm{start}$ &$\lambda_\mathrm{cent}$ & $\lambda_\mathrm{end}$ &
    Exposures & Shift \\
    & [\AA] & [cm$^{-2}$] & [\AA] & [\AA] & [\AA] & [\#] & [km/s] \\
    \hline \\ [-2.0ex]
    MgI  & 2852.96 &    86 & 6717.86 & 6719.73 & 6721.61 & 13-18 & -0.1792\\
    SiII & 1526.71 &    50 & 3594.70 & 3595.65 & 3596.60 & 1-3   & +0.0271\\
    SiII & 1808.01 &   520 & 4257.63 & 4258.55 & 4259.47 & 4-9   & -0.0223\\
    CrII & 2056.27 & -1110 & 4842.33 & 4843.25 & 4844.17 & 4-12  & -0.0059\\
    CrII & 2066.16 & -1360 & 4865.80 & 4866.76 & 4867.71 & 4-12  & -0.0012\\
    MnII & 2576.89 &  1420 & 6068.10 & 6069.50 & 6070.90 & 13-18 & +0.0539\\
    MnII & 2594.51 &  1148 & 6109.60 & 6111.05 & 6112.50 & 13-18 & +0.0622\\
    MnII & 2606.48 &   986 & 6137.50 & 6139.00 & 6140.50 & 13-18 & +0.0678\\
    NiII & 1709.60 &   -20 & 4025.50 & 4026.63 & 4027.76 & 4-9   & -0.0687\\
    NiII & 1741.55 & -1400 & 4100.85 & 4101.94 & 4103.04 & 4-9   & -0.0536\\
    NiII & 1751.92 &  -700 & 4125.40 & 4126.50 & 4127.61 & 4-9   & -0.0487\\
    FeII & 2382.76 &  1460 & 5610.96 & 5612.48 & 5614.00 & 10-12 & -0.0375\\
    FeII & 2600.17 &  1330 & 6122.46 & 6124.22 & 6125.98 & 13-18 & +0.0648\\
    FeII & 2344.21 &  1210 & 5520.17 & 5521.39 & 5522.60 & 10-12 & -0.0557\\
    FeII & 2586.65 &  1490 & 6091.09 & 6092.53 & 6093.98 & 13-18 & +0.0585\\
    FeII & 1608.45 & -1300 & 3787.54 & 3788.51 & 3789.47 & 1-9   & -0.0523\\
    FeII & 2374.46 &  1590 & 5591.45 & 5592.79 & 5594.13 & 10-12 & -0.0414\\
    FeII & 1611.20 &  1100 & 3794.08 & 3794.93 & 3795.77 & 1-9   & -0.0510\\
    \hline \\ [-2.0ex]
  \end{tabular}
  \label{table:reglist}
\end{table*}

\section{Recovering $\Delta\alpha/\alpha$ and $\gamma$}
\label{recovery}

In this section, we show how to recover \da\ and the distortion 
slope $\gamma$ from the distorted simulated spectrum. We also 
demonstrate that fitting the wrong distortion model can lead to 
significant systematic errors on both the \da\ estimates and the 
recovered distortion slope.

\subsection{Analysis}
\label{recover}

We initially apply the distortion illustrated in the middle panel of
Figure \ref{fig:distmodel} 
to the simulated spectrum by applying fixed velocity shifts, determined
from the input distortion model, to each transition.  This is done 
directly via the \vpfit{} input file.  The slope of the distortion model,
$\gamma$, is a free parameter that we solve for.\\

The redshifts of corresponding velocity components in all species 
are tied in the fit, and since the absorption system model is 
turbulent, $b$-parameters of all species are also tied accordingly.  
This procedure is repeated for small increments of 0.005 in $\gamma$ 
over the range +0.1 m/s/\AA\ to +0.25 m/s/\AA. The parabolic relationship
between $\chi^2_\mathrm{min}$ and $\gamma$ is fitted using a third order
polynomial and the relationship between \da\ and $\gamma$ linearly,
enabling us to recover the best-fit values for both parameters,
with associated parameter errors.\\

The whole procedure above is carried out twice, once where we fit 
the sawtooth distortion model (i.e. the same model used to distort
the simulated spectrum) to the data and again but fitting the
simplistic single linear model of \citetalias{whitmore:2014}.

\subsection{Results} \label{results}

\subsubsection{Sawtooth distortion model}

\begin{figure}
  \centering
  \includegraphics[width=\columnwidth]{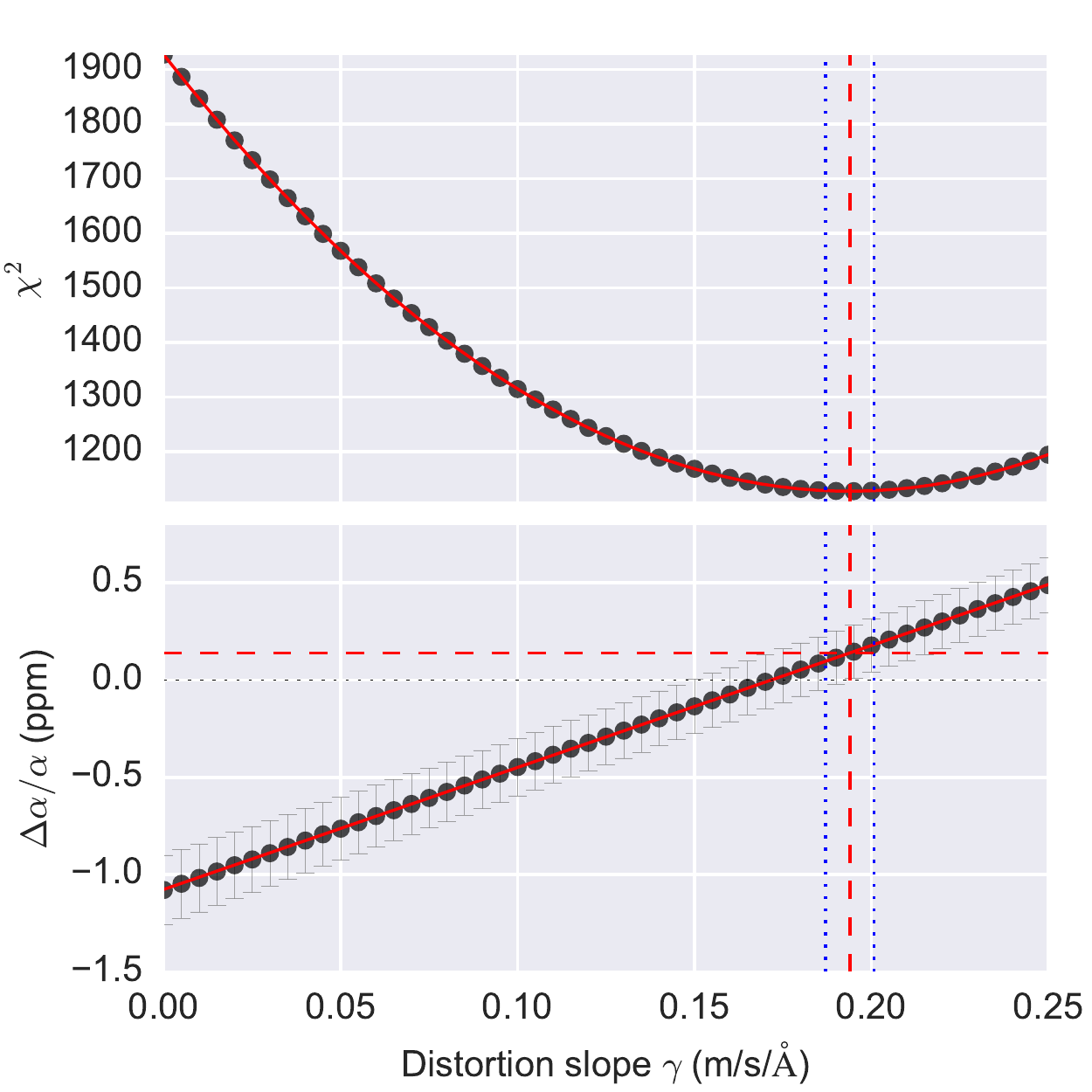}
  \caption{$\chi^2$ and \da\ curves using the sawtooth distort
  ion
  model. The top panel shows the result from fitting the
  simulated spectrum using a sawtooth model, whose properties
  (apart from the slope, which is a free parameter) 
  are determined by knowing the original observational details.
  The bottom panel shows how the inferred \da\ depends on the actual 
  distortion slope.  The best fit gives a distortion slope
  of $0.2062\pm0.0073$ m/s/\AA\ with a corresponding \da\
  estimate of $-0.116\pm0.149$ ppm.}
  \label{fig:rightchisq}
\end{figure}

The results of fitting the simulated spectra using the sawtooth distortion 
model (Figure \ref{fig:distmodel}) are shown in Figure \ref{fig:rightchisq}.  
The uncertainty on $\gamma$ is derived using the standard approach of 
$\chi^2_\mathrm{min} \pm 1$ for one ``interesting'' parameter.  
The uncertainty on \da\ then follows from projection of the parabolic
uncertainty on $\gamma$ as Figure \ref{fig:rightchisq} illustrates.\\

It is worth emphasising that the uncertainties we derive are determined
by the absorption system characteristics used and the high signal to
noise used in these simulations. The uncertainties derived using the
synthetic spectrum described above should therefore not be considered as
representative of existing observational data.\\

After fitting the chi-square curve using a third order polynomial 
and the \da\ curve linearly, we find a best-fit distortion slope of
$\gamma = 0.2062\pm0.0073$ m/s/\AA.  We thus recover the input slope of 
$0.2$ m/s/\AA\ to high precision.\\

The recovered value of \da\ is $-0.116\pm0.149$ ppm and is 
consistent with the null input value.\\

Over the small range in distortion slope considered, the relation
between \da\ and $\gamma$ is, to a good approximation, linear.
However, in general, that need not necessarily be so.  
As the distortion slope, $\gamma$, changes, different transitions shift 
by different amounts, which impacts on the measured $\chi^2$ for the best-fit.
If we consider a large range in distortion slope, non-linearities
begin to appear, and the velocity structure in the model can then even
change, causing discontinuities in \da\ vs $\gamma$.  We shall discuss
this issue in a separate paper, although we can say here that it is
generally not an important problem because $\gamma$ is usually well-constrained
by the quasar spectrum itself to lie within a small range, provided 
there is reasonable set of transitions with a good range in q-coefficients
contributing to the fit.

\subsubsection{Simplistic distortion model}

\begin{figure}
  \centering
  \includegraphics[width=\columnwidth]{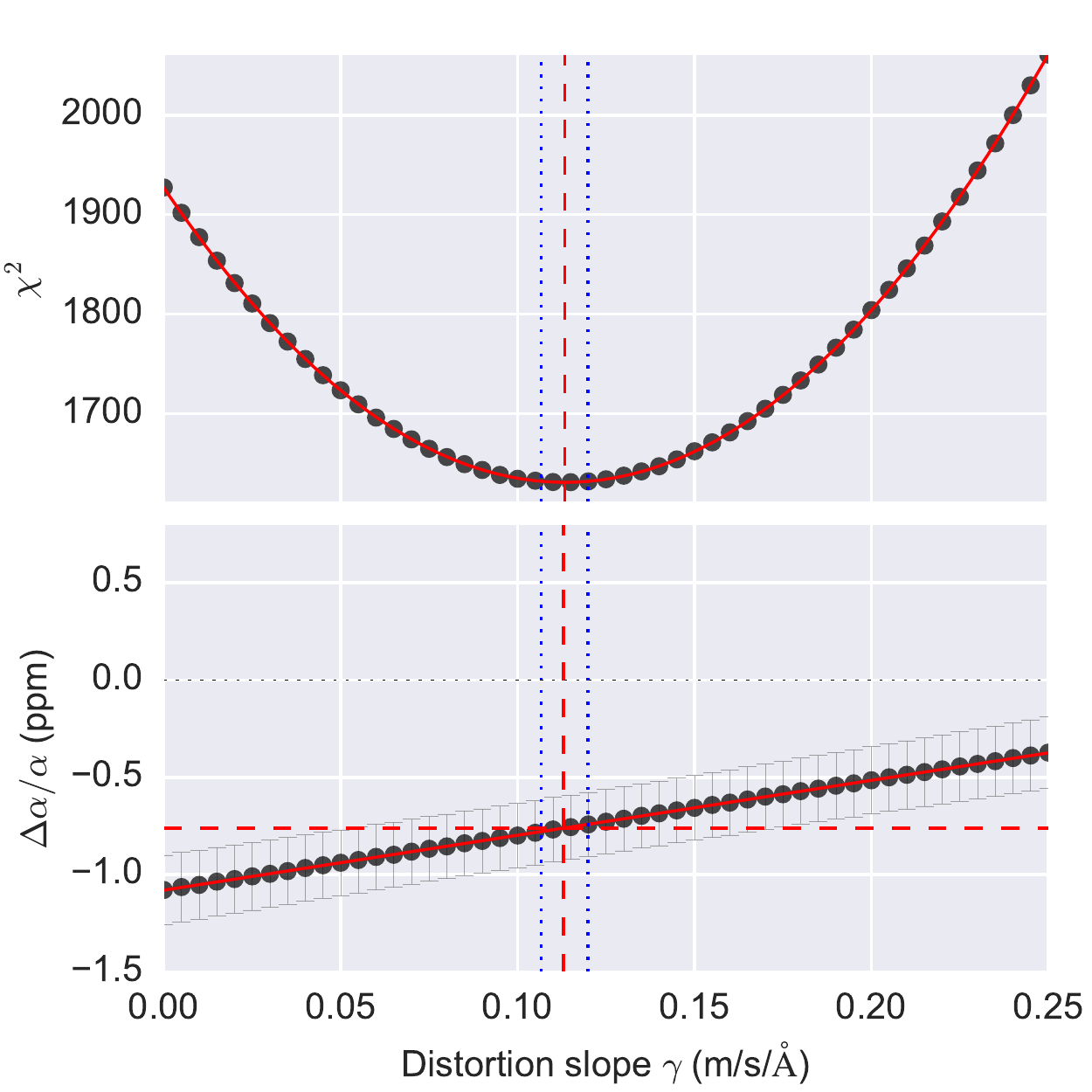}
  \caption{$\chi^2$ and \da\ curves using the simplistic distortion
  model of \citetalias{whitmore:2014}.  The top panel shows the best-fit
  of $0.1324\pm0.0068$ 
  m/s/\AA, 10$\sigma$ away from the true (input) value of $\gamma=0.2$.
  The bottom panel illustrates the ``corrected'' \da\ of 
  $-0.771\pm0.173$ ppm, a 4.5$\sigma$ departure from the true (input) 
  value of \da$=0$.}
  \label{fig:wrongchisq}
\end{figure}

The results of fitting the simulated spectra using the simplistic linear  
distortion model (bottom panel of Figure \ref{fig:distmodel}) are shown in
Figure \ref{fig:wrongchisq}.  After again fitting the chi-square curve using 
a third order polynomial and the \da\ curve linearly, we find a best-fit 
distortion slope of $\gamma = 0.1324 \pm 0.0068$ m/s/\AA, inconsistent at 
the 10$\sigma$ level with the correct value (i.e. the value used to create 
the simulated spectrum) of $0.2$.\\

The recovered value of \da\ is $-0.771\pm0.173$ ppm,  inconsistent with 
the null input value at the 4.5$\sigma$ level.

\subsubsection{Comparison between models}

It is interesting to note that a decent parabolic shape for
$\chi^2$ vs $\gamma$ is obtained for both models.  However,
the inferred corrections to \da\ are quite different.  The 
correction on \da\ inferred from the WM model shifts the result 
from the ``uncorrected'' value (corresponding to $\gamma=0$ in 
Figure \ref{fig:distmodel}) from 
$-1.36$ ppm to $-0.77$ ppm, i.e. a correction shift of $+0.59$ ppm.
However, the corresponding shift derived from the sawtooth model 
is from $-1.36$ ppm to $-0.12$ ppm i.e. a correction shift of 
$+1.24$ ppm, more than twice as large.  The latter is consistent 
with the true input value of zero.\\

We define the \da\ correction shift as the difference between the \da\
results at the best-fit distortion slope and a slope of 0. When no
distortion slope is applied to the model, the resulting \da\ is
found to be consistent between the two approaches, with a
best-fit value of  $\Delta\alpha/\alpha=-1.306\pm0.184$. 

The \da\ correction derived using the sawtooth model is more than twice
as large as the \da\ correction derived using the simplistic model. The
sawtooth distortion model gives a \da\ correction shift of
$-0.116-(-1.306)=1.19$, the resulting shift when the simplistic
distortion model is applied corresponds to $-0.771-(-1.306)=0.535$.
These shifts and their uncertainties are absorption system dependent and
hence need to be solved for and applied on a system-by-system basis.
Whether or not these corrections translate into any kind of redshift
dependence must be determined for any statistical sample.

\section{Conclusions}

We have investigated long-range wavelength distortions in echelle
spectrographs, in the context of quasar spectroscopy, in order to
quantify the impact on measurements of \da.  We created realistic
numerical simulations of a known absorption system at
$z_\mathrm{abs}=1.3554$ towards the $z_\mathrm{em}=1.94$ quasar
J043037-485523 that has previously been used to measure \da.  Long-range
wavelength distortions, based on observations of asteroids and
solar-twins with the UVES instrument on the VLT, were imposed on the
simulated spectra.  The simulated spectra were then fitted in two ways:
first using the same distortion model but treating the slope of the
distortion relation as a free parameters, and second with a simplistic
distortion model used previously in \citetalias{whitmore:2014}. The main
results are:

\begin{itemize}
\item[1.] If the simplistic distortion model of
\citetalias{whitmore:2014} is used to solve for long-range distortion,
and hence to correct \da\ measurements, a significant systematic offset
in \da\ is introduced, emulating non-zero results.  The correction done
in this way does not work.\\

\item[2.] For the one specific absorption model we have considered, the
systematic offset in \da\ introduced by using an incorrect distortion
model is substantially greater than typical statistical measurement
errors. This suggests that the inference of \citetalias{whitmore:2014}
that long-range wavelength distortions may account for the spatial dipole
reported in \citet{webb:2011} and \citet{king:2012} is unlikely to be
correct.\\

\item[3.] If instead a more appropriate distortion model is used, 
allowing for the way in which almost all quasar spectra have previously 
been observed (using multiple wavelength settings for multiple 
exposures), no systematic \da\ offsets are found and the long-range 
distortion corrections to \da\ accurately recover the true \da.
\end{itemize}

Clearly the quantitative results given above (i.e. the statistical
significances) are model-dependent, since we simulated one particular 
quasar absorption system to illustrate the results. Nevertheless, the
generality of the conclusions above are supported by applying the
method described in this paper to the large UVES sample of \da\
measurements used in \citet{webb:2011} and \citet{king:2012}. That
work will be reported in a separate paper.\\

It is not our aim in this paper to claim that the spatial dipole 
reported in \citet{webb:2011} and \citet{king:2012} is correct.
Whether that is so is still to  be determined using larger statistical
samples, and/or independent methods, and/or by the discovery of
some systematic that explains it. However, we do wish to emphasise
that no  systematic effect has yet been found that emulates the
spatial dipole.

\nocite{*}
\bibliography{papers}

\label{lastpage}

\end{document}